# On the Reproducibility of TCGA Ovarian Cancer MicroRNA Profiles

Ying-Wooi Wan[1,2,4], Claire M. Mach[2,3], Genevera I. Allen[1,7,8], Matthew L. Anderson[2,4,5]*, Zhandong Liu[1,5,6,7]*

[1]Departments of Pediatrics [2]Obstetrics and Gynecology and [4]Pathology and Immunology, [5]Dan L. Duncan Cancer Center, [6]Computational and Integrative Biomedical Research (CIBR) Center, Baylor College of Medicine.

[3]College of Pharmacy, University of Houston

[7]Neurological Research Institute, Texas Children's Hospital

[8]Department of Statistics and Electrical & Computer Engineering, Rice University

*<u>Correspondence</u>:

*One Baylor Plaza, BCM320*

*Phone: 832-824-8788*

*FAX: 713-789-4855*

*Email: Zhandonl@bcm.edu or matthew@bcm.edu*

**ABSTRACT**

Dysregulated microRNA (miRNA) expression is a well-established feature of human cancer. However, the role of specific miRNAs in determining cancer outcomes remains unclear. Using Level 3 expression data from the Cancer Genome Atlas (TCGA), we identified 61 miRNAs that are associated with overall survival in 469 ovarian cancers profiled by microarray ($p<0.01$). We also identified 12 miRNAs that are associated with survival when miRNAs were profiled in the same specimens using Next Generation Sequencing (miRNA-Seq) ($p<0.01$). Surprisingly, only 1 miRNA transcript is associated with ovarian cancer survival in both datasets. Our analyses indicate that this discrepancy is due to the fact that miRNA levels reported by the two platforms correlate poorly, even after correcting for potential issues inherent to signal detection algorithms. Further investigation is warranted.

**INTRODUCTION**

MicroRNAs (miRNAs) are endogenous RNA transcripts that regulate diverse patterns of gene expression.[1] Most human miRNAs are transcribed as long precursors known as pri-miRNAs. Starting in the nucleus, pri-miRNAs undergo a series of processing events that ultimately result in the cytoplasmic release of mature transcripts ~22 nucleotides in length. Mature miRNAs catalyze translational inhibition by directly binding to messenger RNAs (mRNAs) and promoting their degradation.[2] Recent data also indicate that miRNAs can inhibit translation independent of their ability to induce mRNA degradation.

Patterns of miRNA expression have now been extensively profiled in many different human tissues. It is now clear that dysregulated miRNA expression is a feature of many different cancers, including carcinomas of the breast, ovary and lung.[3-5] However, determining the mechanisms by which individual miRNAs contribute to cancer outcomes remains a key challenge for biologists hoping to exploit their power. Recently, the Cancer Genome Atlas Consortium (TCGA) reported that ovarian cancers cluster into distinct molecular subtypes based on their patterns of gene and microRNA expression.[6] However, we have discovered an alarming lack of consistency between the microRNA (miRNA) expression profiles initially used by the TCGA and a subsequent profile of miRNA expression generated by this group for the same ovarian cancer specimens using miRNA-Seq. As these observations challenge the validity of the underlying data, they also challenge scientific discoveries based on this data.

**RESULTS**

To delineate miRNAs associated with ovarian cancer patient survival, we performed a univariate Cox regression analysis using Level 3 TCGA miRNA data for 469 ovarian cancers profiled using Agilent microarray technology. We found that 61 mature miRNAs are significantly associated with ovarian cancer survival (p<0.01) (Figure 1A). Of these, miR-505, miR-652 and miR-551b* demonstrate the most robust associations. Hazard ratios (HR) calculated for these miRNAs were -1.73, -1.8, and 9.3, respectively, indicating that each miRNA potentially plays an important role in determining ovarian cancer survival.

To validate these observations, we next interrogated a second dataset of miRNA expression generated for the same ovarian cancer specimens using Next Generation Sequencing (miRNA-Seq). The TCGA ovarian cancer project is unique in that miRNA expression has been profiled using both miRNA array and miRNA-Seq. Use of these technically distinct platforms creates a unique opportunity to validate discoveries made using one dataset against the other. Ideally, the results obtained should correlate tightly. Using Cox Proportional Hazards analysis, we found that 12 miRNA transcripts are associated with survival when miRNAs were profiled in ovarian cancers using miRNA-Seq (Figure 1B). However, the hazard ratios estimated for the 12 miRNAs identified from miRNA-Seq data are all very close to 1.0. Surprisingly, only miR-652 is associated with survival in both the miRNA-Seq and microarray datasets. To correct for multiple hypothesis testing, we adjusted our Cox model p-values using Benjamini–Hochberg procedure.[7] After completing these analyses, no miRNAs are correlated with survival in both datasets when the false discovery rate was set at 10%.

To elucidate potential causes for this unexpected discrepancy, we examined the reproducibility of miRNA expression between the two TCGA files that describe this data. Pearson correlation coefficients (r) were calculated for each of the 359 mature human miRNAs for which Level 3 expression data was available in both the miRNA-Seq and microarray databases. We found that correlation coefficients for levels of individual miRNAs reported by each technique varied widely. For example, miR-505 is the miRNA most robustly associated with patient outcome in our analyses of the miRNA array data (HR = -1.7, p< 9e-5). However, when assessed using sequencing data, the hazard ratio for mir-505 was 0.998 (p=0.03). Levels of miR-505 measured by miRNA-array and miRNA-Seq data correlated only modestly (r = 0.59) (Figure 2B). Discrepancies were also observed in a number of other miRNAs that have been previously implicated in ovarian cancer, such as miR-143.[8] The correlation coefficient for miR-143 in our analyses was 0.39 (Figure 2C). Another miRNA well-studied in ovarian cancer is miR-141, which has been previously reported to target p38α and modulate the oxidative stress response.[9,10] However, the correlation between levels of miR-141 in TCGA microarray and miRNA-Seq expression data is only 0.32 (Figure 2D). Overall, we found that correlation coefficients for ~72% of miRNAs profiled in both datasets were ≤ 0.5 (Figure 3A, 3C), indicating poor reproducibility. In contrast, only 22% of the mRNAs measured by Agilent microarray and Illumina HiSeq using the same ovarian cancer specimens correlate poorly (r ≤ 0.5; Figure 3B, 3C). Thus, the discrepancy we report here appears to be limited to the TCGA miRNA dataset.

One potential cause for poor reproducibility may be the signal detection algorithm used to report levels of miRNA expression. Level 3 TCGA miRNA data are reported in two formats. The first, labeled as a "Quantification Data," reports levels for individual human miRNAs. However, one of the advantages of miRNA-Seq is that transcripts retrieved by this technique can be precisely mapped. A second file, labeled as "Isoform Data," has also been released by the TCGA. This file reports read counts for transcripts according to their genomic location. As part of this file, transcripts are identified as either mature miRNA, miRNA* (3p arms of human miRNAs), stem-loop transcript or precursor. While working through this data, we learned that miRNA levels reported in the TCGA quantification file include read counts for miRNA precursors as well as mature miRNAs. Because miRNA precursors typically lack biologic activity, inclusion of precursors with counts for mature miRNAs could confound survival analyses. To address this issue, we retrieved read counts for mature miRNAs only from the isoform data file and repeated our analyses. However, the proportion of miRNA correlation coefficients ≤ 0.5 remained as high as 71% despite the use of this more precisely defined data.

A second possible explanation for the discrepancy we observed might be that correlations between measures of miRNA expression depend on the frequency with which individual miRNA transcripts are expressed. If so, infrequently expressed miRNAs might be reported by one or both of the platforms used to profile miRNA expression randomly or inaccurately. To explore this hypothesis, we re-calculated correlation coefficients for each miRNA identified by both platforms after excluding any transcript in the miRNA-Seq dataset with a read count less than 5. This reduced the number of distinct miRNAs available for analysis in the miRNA-Seq data file from 705 to 380. However, the proportion of miRNAs with correlation coefficients ≤ 0.5 also decreased from 72% to 56%. Similarly removing poorly expressed transcripts from the pool of mRNAs profiled by Illumina HiSeq reduces the proportion of mRNAs whose correlation coefficients ≤ 0.5 from 22% to 20%. These observations indicate that problems detecting infrequently expressed miRNA may impact the ability or one or both platforms to reliably report miRNA expression. However, the fact that more than half of miRNA transcripts still had correlation coefficients ≤ 0.5 even after correcting for this issue indicates that poorly expressed transcripts are not solely responsible for the discordant patterns of miRNA expression reported by the two platforms.

**DISCUSSION**

Much to our surprise, our analyses indicate that the microRNAs associated with survival in ovarian cancer depend highly on whether specimens were profiled by the TCGA using microarray or miRNA-Seq. Our analyses indicate that this discrepancy exists because miRNA-Seq and microarray have generated very different profiles of miRNA expression, even though the data is based on the same ovarian cancer specimens. We do not currently have a clear explanation for why miRNA expression profiles reported by the TCGA are discordant. However, understanding this discrepancy will ultimately be important for identifying which miRNAs if any are important for determining ovarian cancer outcomes.

A variety of DNA microarray technologies have been previously validated by investigators examining within platform and cross-platform reproducibility.[11-13] Spearman correlation coefficients reported in these studies range from 0.59 to 0.94 with a mean of 0.82. These results are similar to what we have observed for correlations between patterns of gene expression profiled using microarray and Illumina HiSeq platforms by the TCGA. Both miRNA-Seq and microarray technologies are associated with multiple technical limitations that might account for the differences we have observed. For example, cross-hybridization is a well-recognized issue that can reduce signal specificity when profiling RNA transcripts by microarray.[14] However, it seems unlikely that cross-hybridization can fully explain the discrepancy we observed, as the number of transcripts correlated with survival by array is greater than the number associated with survival by miRNA-Seq. One alternate explanation might be that the signal extraction algorithm used to analyze miRNA-Seq data does not accurately report miRNA levels. In general, miRNA-Seq allows for precise transcript mapping with much greater confidence. The signal extraction algorithm currently used by the TCGA to report miRNA levels includes read counts for both a mature miRNA and its corresponding precursor. Precursors account for fewer than 1% of the total miRNA counts in the TCGA isoform file, likely reflecting the use of size-fractionated total RNA to prepare small RNA libraries for miRNA-Seq.[5] Our analyses indicate that their inclusion has little bearing on which miRNAs are associated with ovarian cancer survival.

These observations underscore the urgent need for well-defined algorithms for processing signals generated by miRNA-Seq and transcriptional profiling platforms. Our understanding is that the same analyses have been performed by TCGA for other cancers, including colon, breast and lung.[15-17] Because miRNA expression in these other cancers has not been profiled by microarray, it is not possible to repeat our analyses to determine whether the discrepancy we report is observed in other cancers.

Ultimately, consistent and reliable genomic data is critical for constructing testable hypotheses and achieving the full potential of the TCGA. Our observations identify an important hazard of which investigators should be aware as they utilize the TCGA miRNA data to study ovarian cancer. This hazard underscores the need to validate observations made with one or both of TCGA miRNA datasets. Over the long term, resolution of this discrepancy will be important for determining the most effective platform and signal extraction algorithms for profiling miRNA expression as part of large scale genomic profiling efforts.

**MATERIALS AND METHODS**

**Gene and microRNA Expression Data.** Level 3 data documenting patterns of gene expression for 296 ovarian cancer specimens profiled using Agilent G4502A arrays and Illumina HiSeq were downloaded from the TCGA data portal. Level 3 microRNA expression data were also retrieved for 469 ovarian cancer specimens profiled using the Agilent 4X15k array and miRNA-Seq. Level 3 miRNA data profiled by miRNA-Seq were retrieved from both the miRNA quantification and isoform files available at the TCGA data portal along with metafiles annotating each dataset. Permission to access all data was obtained from the Data Access Committee for the National Center for Biotechnology Information Genotypes and Phenotypes Database (dbGAP) at the National Institutes of Health.

**Survival Analyses.** Coded patient survival data was extracted from the TCGA clinical information file. A Cox Proportional Hazards model was used to estimate association between levels of individual miRNAs. Patient survival was calculated as time in months elapsed from date of diagnosis until date of last contact.

**Statistical Analyses.** Spearman's rank correlation coefficients, histograms, and the empirical cumulative distribution were computed and plotted for each miRNA and gene using r. Sequencing data were log transformed for plotting. Both direct read counts and counts normalized according to millions of miRNAs were examined as part of our analyses. All analyses were performed using both raw and normalized read counts reported as part of the TCGA miRNA-Seq datasets.

**ACKNOWLEDGEMENTS**

The authors gratefully acknowledge communication from David Wheeler, Rehan Akban, Gordon Robertson and Andy Chu regarding TCGA miRNA data analysis algorithms.## REFERENCES

1. Wu, L. & Belasco, J. G. Let me count the ways: mechanisms of gene regulation by miRNAs and siRNAs. *Molecular cell* **29**, 1-7, doi:S1097-2765(07)00881-7 [pii] 10.1016/j.molcel.2007.12.010 (2008).

2. Bartel, D. P. MicroRNAs: target recognition and regulatory functions. *Cell* **136**, 215-233, doi:10.1016/j.cell.2009.01.002 (2009).

3. Eder, M. & Scherr, M. MicroRNA and lung cancer. *The New England journal of medicine* **352**, 2446-2448, doi:10.1056/NEJMcibr051201 (2005).

4. Zhang, L. *et al.* microRNAs exhibit high frequency genomic alterations in human cancer. *Proc Natl Acad Sci U S A* **103**, 9136-9141 (2006).

5. Creighton, C. J. *et al.* Molecular profiling uncovers a p53-associated role for microRNA-31 in inhibiting the proliferation of serous ovarian carcinomas and other cancers. *Cancer Res* **70**, 1906-1915, doi:0008-5472.CAN-09-3875 [pii] 10.1158/0008-5472.CAN-09-3875 (2010).

6. Integrated genomic analyses of ovarian carcinoma. *Nature* **474**, 609-615, doi:10.1038/nature10166 (2011).

7. Benjamini, Y. & Hochberg, Y. Controlling the False Discovery Rate: A Practical and Powerful Approach to Multiple Testing. *Journal of the Royal Statistical Society. Series B (Methodological)* **57**, 289-300, doi:10.2307/2346101 (1995).

8. Marchini, S. *et al.* Association between miR-200c and the survival of patients with stage I epithelial ovarian cancer: a retrospective study of two independent tumour tissue collections. *Lancet Oncol* **12**, 273-285, doi:10.1016/S1470-2045(11)70012-2 (2011).

9. Mateescu, B. *et al.* miR-141 and miR-200a act on ovarian tumorigenesis by controlling oxidative stress response. *Nature Medicine* **17**, 1627-1635, doi:10.1038/nm.2512 (2011).

10. Yang, D. *et al.* Integrated Analyses Identify a Master MicroRNA Regulatory Network for the Mesenchymal Subtype in Serous Ovarian Cancer. *Cancer Cell* **23**, 186-199, doi:10.1016/j.ccr.2012.12.020 (2013).

11. Sato, F., Tsuchiya, S., Terasawa, K. & Tsujimoto, G. Intra-platform repeatability and inter-platform comparability of microRNA microarray technology. *PloS one* **4**, e5540, doi:10.1371/journal.pone.0005540 (2009).

12. Yauk, C. L., Rowan-Carroll, A., Stead, J. D. & Williams, A. Cross-platform analysis of global microRNA expression technologies. *BMC genomics* **11**, 330, doi:10.1186/1471-2164-11-330 (2010).

13. Shi, L. *et al.* The MicroArray Quality Control (MAQC) project shows inter- and intraplatform reproducibility of gene expression measurements. *Nature Biotechnology* **24**, 1151-1161, doi:10.1038/nbt1239 (2006).

14. Wu, C., Carta, R. & Zhang, L. Sequence dependence of cross-hybridization on short oligo microarrays. *Nucleic Acids Research* **33**, e84, doi:10.1093/nar/gni082 (2005).

15. Comprehensive molecular characterization of human colon and rectal cancer. *Nature* **487**, 330-337, doi:10.1038/nature11252 (2012).

16. Comprehensive genomic characterization of squamous cell lung cancers. *Nature* **489**, 519-525, doi:10.1038/nature11404 (2012).
9

**FIGURES**

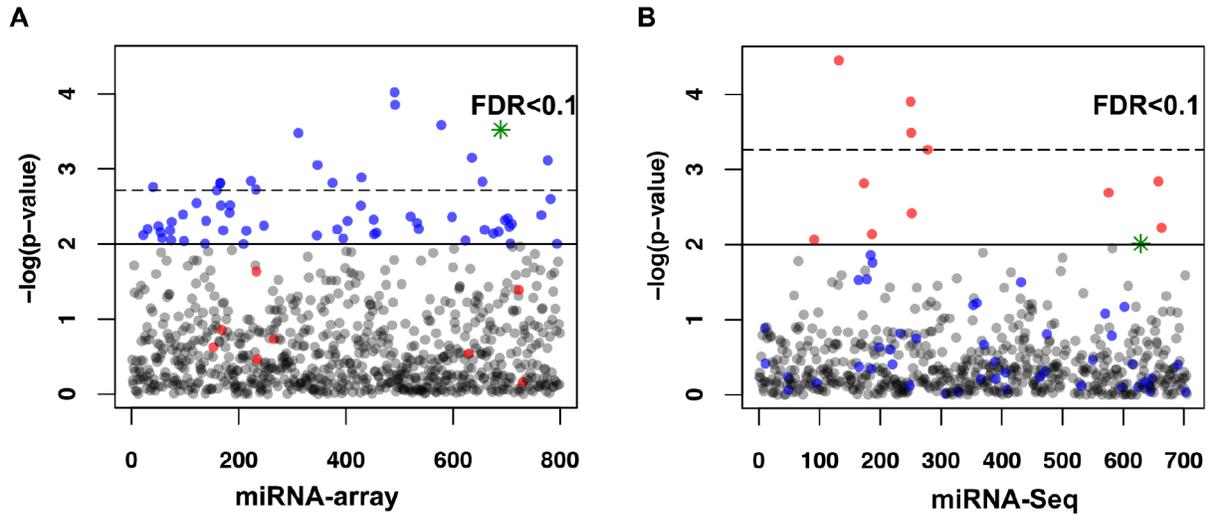

**FIGURE 1. MicroRNAs associated with ovarian cancer survival.** P-value plots of univariate Cox regression for microRNAs associated with ovarian cancer survival identified by microarray (A) or miRNA-Seq (B) data. P-value < 0.01 (Solid line). False discovery rate (FDR) < 0.1 (Dotted line). In both A&B, blue dots indicate miRNAs associated with survival by miRNA array, while red dots indicate miRNAs associated with survival by miR-Seq. Green stars are miRNAs associated with survival in both datasets.

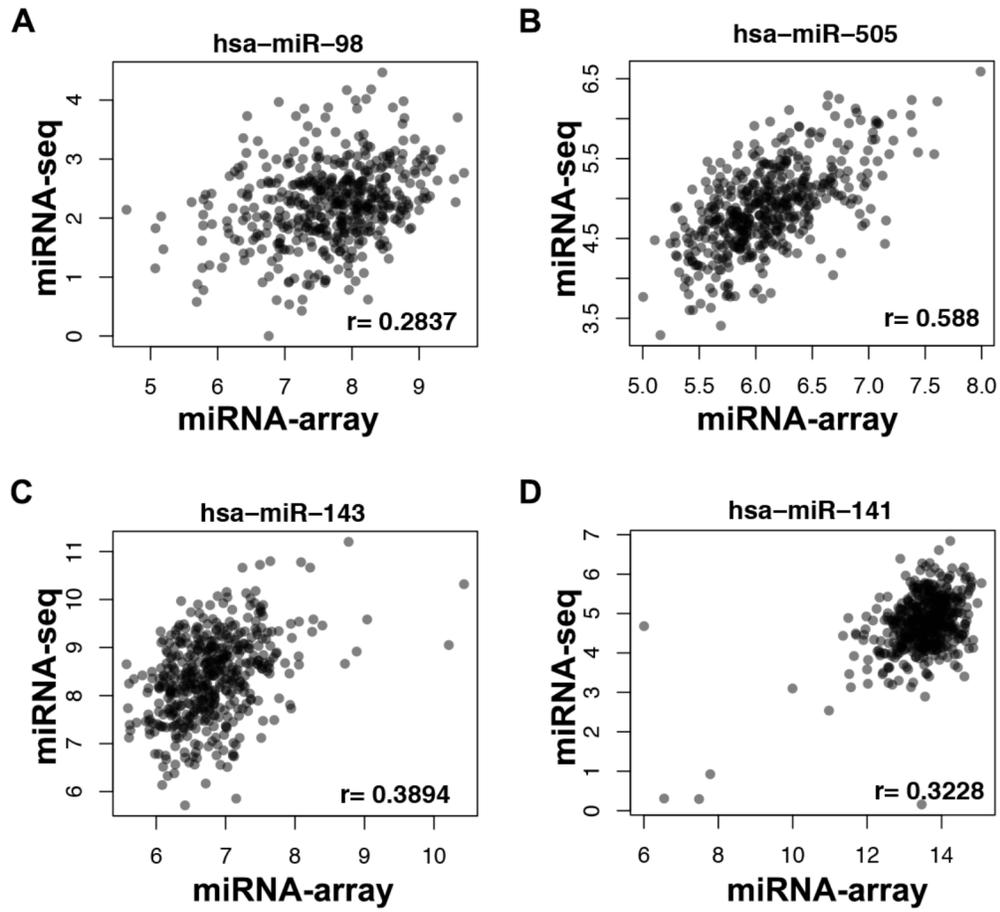

**FIGURE 2. Scatter-plots of microRNA expression measured by microarray and miRNA-Seq.** (A) miR-98, (B) miR-505 (C) miR-143 and (D) miR-141.

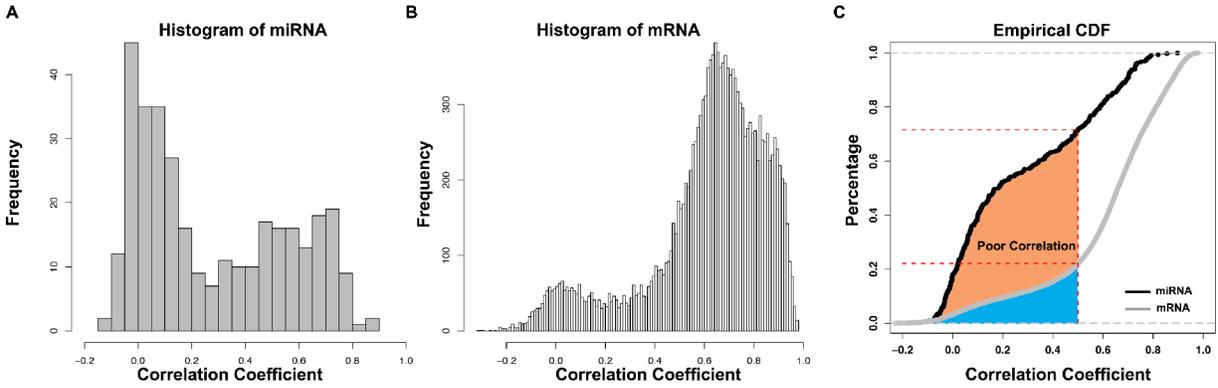

**FIGURE 3. Distribution of correlations between microarray and sequencing profiles for miRNA and gene expression**. (A) Histogram of correlation coefficients for individual miRNAs measured by miRNA-Seq and miRNA array. (B) Histogram of correlation coefficients for mRNAs profiled by Illumina HiSeq and mRNA array. (C) The empirical cumulative distribution function (ECDF) of the correlation between array and sequencing for miRNA (black) and mRNA (gray) measurements. Nearly, 72% of miRNAs demonstrate a correlation coefficient ≤ 0.5 whereas 22% of RNAs have a correlation coefficient ≤ 0.5.